\documentclass[fleqn,twoside]{article}
\usepackage[headings]{espcrc2}

\usepackage{graphicx}
\usepackage[figuresright]{rotating}

\title{Status of PHOKHARA
\thanks{Work supported in part by 
EU 6th Framework Programme under contract MRTN-CT-2006-035482 
(FLAVIAnet).}}

\author{     Agnieszka Grzeli\'nska \address{Institute of Nuclear Physics 
Polish Academy of Sciences, PL-31342 Cracow, Poland},
Henryk Czy\.z \address[MCSD]{Institute of Physics, University of 
Silesia, PL-40007 Katowice, Poland} %
        and
    Agnieszka Wapienik \addressmark[MCSD] }
       
\begin{document}

\begin{abstract}
A review of the status of the Monte Carlo event generator PHOKHARA,
developed for experiments using the radiative return method.
The four-pion production in electron-positron annihilation
and in $\tau$-lepton decays  and the narrow resonances 
 studies are described.
%\vspace{1pc}
\end{abstract}


\maketitle

% INTRODUCTION

\section{Introduction}

The basis of the radiative return method, proposed in \cite{Zerwas},
is an observation that one can extract the hadronic cross section
($e^+e^- \to hadrons$) from the measurement of the cross section
of the reaction $e^+e^- \to hadrons + photons$, where
  the photons are  emitted from the initial
leptons. This is possible due to the factorization $d\sigma(e^+e^-\to
hadrons + \gamma_{ISR}) = H(Q^2,\theta_\gamma) d\sigma(e^+e^- \to hadrons)(s=Q^2)$,
where the function $H$ is fully calculable within QED and $Q^2$ is the 
invariant mass of the hadronic system. 

The radiative return method  has been used by meson factories 
DAPHNE, BaBar and BELLE and
allows for the measurement of the hadronic cross
section from the nominal energy of these experiments down to the production 
threshold.

The traditional way of measuring of the hadronic cross section via 
the energy scan needs dedicated experiments, 
 while using the radiative return one can profit from the existing
  meson factories.
The smaller cross section of the radiative process 
(by a  factor of $\alpha / \pi$ 
as compared to the process without photons emission) 
has to be compensated by higher luminosities of factories. 

To obtain the hadronic cross section using the radiative
return method in a realistic experimental situation, 
one needs a Monte Carlo event generator of the measured process. 
To provide such a tool to the experimental groups
 PHOKHARA event generator was constructed.
The construction of the PHOKHARA event generator started from 
the EVA \cite{Binner:1999bt,Czyz:2000wh}
 generator, where structure function method
was used to model multi-photon emission. The physical accuracy of
the program was however far from the demanding experimental accuracy.
To meet this ever growing demands, the event generator PHOKHARA 
\cite{Rodrigo:2001kf} was constructed.
It is  based on the 
complete calculation of radiative corrections to the next-to-leading order
 for the ISR emission and relevant next-to-leading order corrections
 to the final state emission. 
In this paper the latest updates of  the four-pion channels 
\cite{4pi} are briefly outlined (Section 2) and preliminary
results for the narrow resonances implementation are 
 presented (Section 3).

% SECTION

\section{The four-pion production in $\tau$ decays and e$^+$e$^-$ annihilation}

The four-pion production in $e^+e^-$ annihilation was implemented in 
the generator EVA \cite{Czyz:2000wh} a long time ago and recently
it was reanalyzed in \cite{4pi}.

There are altogether four different channels accessible  for 4$\pi$ 
production:

$e^+e^-\to  2\pi^+2\pi^-,$

$e^+e^-\to 2\pi^0\pi^+\pi^-$ (a),

$\tau^-\to\nu_\tau 2\pi^-\pi^+\pi^0$ (b), 

$\tau^-\to\nu_\tau 3\pi^0\pi^-$.

Assuming isospin symmetry, the amplitudes of either
(a) \cite{kuhn:4pi} or (b) \cite{Ecker} 
 are sufficient to determine all four amplitudes. 

In EVA and PHOKHARA a choice of \cite{kuhn:4pi} was adopted
 and a function $J_\mu$ (symmetric
under the interchange of $p_1$ and $p_2$
 and antisymmetric under the interchange of $p^+$ and $p^-$)
 is used to model the two $e^+e^-$ four pion channels:
\begin{eqnarray}
\langle \pi^+ \pi^- \pi_1^0 \pi_2^0 | J^3_{\mu} | 0 \rangle \kern+11pt =
\kern+7pt J_{\mu}  (p_1,p_2,p^+,p^-), 
\label{eq1}
\end{eqnarray}
The other matrix elements:
$\langle \pi^+_1 \pi^+_2 \pi^-_1 \pi^-_2 | J^3_{\mu} | 0 \rangle $,
$\langle \pi^- \pi^0_1 \pi^0_2 \pi^0_3 | J^{-}_{\mu} | 0 \rangle $ and
$\langle \pi^-_1 \pi^-_2 \pi^+ \pi^0 | J^{-}_{\mu} | 0 \rangle $
 can be expressed as sum of  $J_{\mu}$ functions with permuted arguments
 \cite{kuhn:4pi}. 

For $e^+e^-$ annihilation the current $J_\mu$ contains the complete information
about the hadronic cross section through
\begin{eqnarray}
 &&\kern-30pt \int \ J^{em}_\mu (J^{em}_\nu)^*  \ \ d\bar\Phi_n(Q;q_1,\dots,q_n)
 \nonumber \\  &=&  
 \frac{1}{6\pi} \left(Q_{\mu}Q_{\nu}-g_{\mu\nu}Q^2\right) \ R(Q^2) \ ,
\label{rr}
\end{eqnarray}
where $R(Q^2)=\sigma(e^+e^-\to hadrons)(Q^2)/\sigma_{point}$.

Similarly for $\tau$ decay we have
\begin{eqnarray}
 &&\kern-30pt
 \int \ J_\mu^{-} J^{-*}_\nu  \ \ d\bar\Phi_n(Q;q_1,\dots,q_n)  \nonumber \\
 &=&\frac{1}{3\pi} \left(Q_{\mu}Q_{\nu}-g_{\mu\nu}Q^2\right) \ R^{\tau}(Q^2) \ 
\label{rrt}
\end{eqnarray}
where $R^\tau$  can obtained from the differential $\tau$ decay rates.

 From the isospin relations between the matrix elements 
 of the four pion hadronic current one obtains
  the relations between $\tau$ decay rates 
and $e^+e^-$ annihilation cross sections
\begin{eqnarray}
 R^{\tau}\left(-\ 0 \ 0 \ 0\right) &=&  \frac{1}{2}\  R\left(+\ + \ - \ -\right)
\nonumber \\
 R^{\tau}\left(-\ - \ + \ 0\right) 
 &=& \frac{1}{2} \ R\left(+\ + \ - \ -\right) \ \nonumber \\
 &+& \ R\left(+\ - \ 0 \ 0\right) \ .
\label{CVC}
\end{eqnarray}

They allow for direct tests of the isospin symmetry 
 provided all $R$ functions were measured.

%%%%%%%%%%%%%%%%%%%%%%%%%%  FIGURE 
 \begin{figure}
\includegraphics[width=7.5cm,height=6cm]{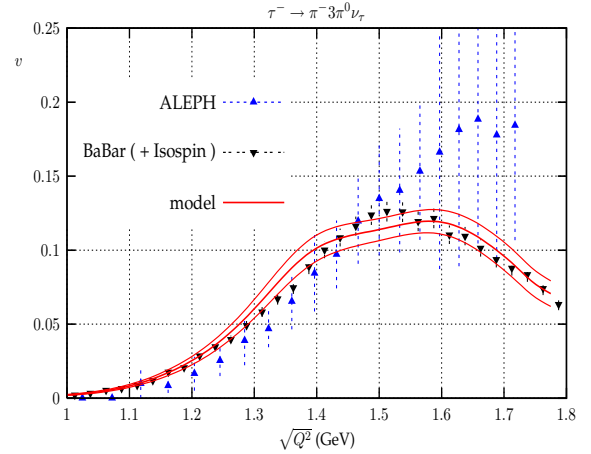}
\vskip -1. cm
\caption{The spectral function of  the $\tau^- \to 3\pi^0 \pi^- \nu_\tau $ 
 decay mode.
 ALEPH \cite{Schael:2005am} data versus 
predictions from BaBar \cite{Aubert:2005eg,bb1} 
and the model \cite{4pi}.}
\label{fun11}
\end{figure}
%%%%%%%%%%%%%%%%%%%%%%%%%

%%%%%%%%%%%%%%%%%%%%%%%%%%  FIGURE 
 \begin{figure}
\includegraphics[width=7.5cm,height=6cm]{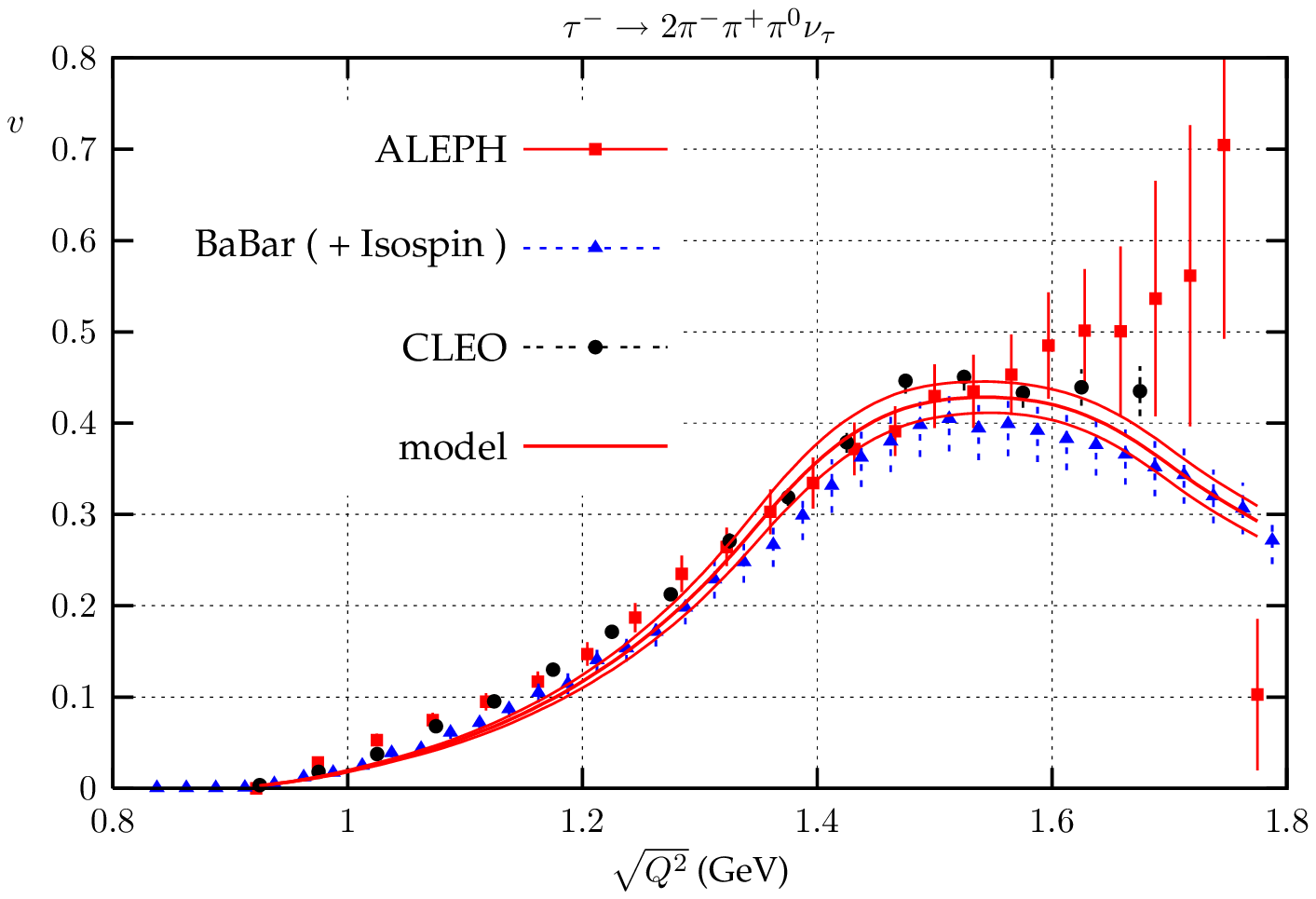}
\vskip -1. cm
\caption{The spectral function of the $\tau^- \to 2\pi^- \pi^+\pi^0 \nu_\tau $
 decay mode.
 ALEPH \cite{Schael:2005am} and CLEO \cite{Edwards:1999fj} data versus 
predictions from BaBar \cite{Aubert:2005eg,bb1} 
and the model \cite{4pi}.}
\label{fun12}
\end{figure}
%%%%%%%%%%%%%%%%%%%%%%%%%

%%%%%%%%%%%%%%%%%%%%%%%%%% 
 \begin{figure}[ht]
\begin{center}
\includegraphics[width=8cm,height=4cm]{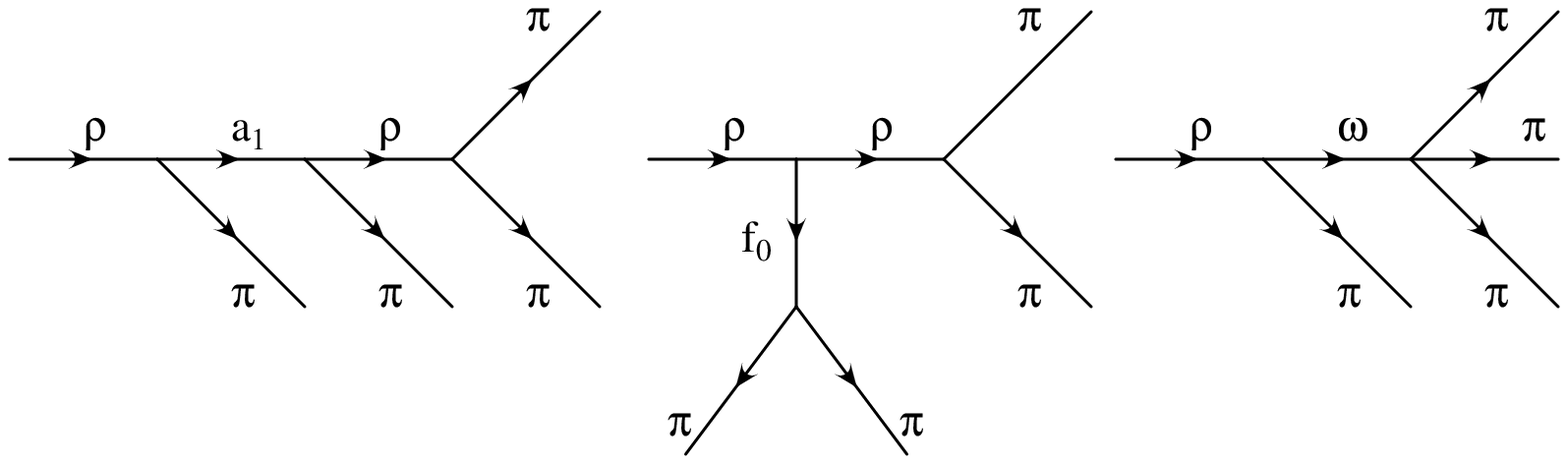}
\includegraphics[width=4cm,height=3.5cm]{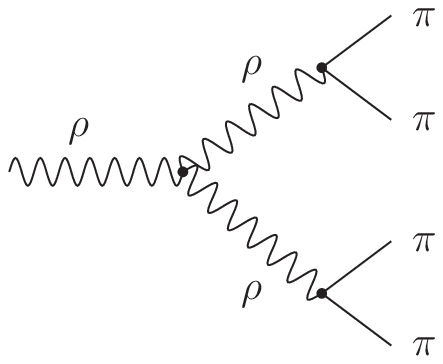}\includegraphics[width=4cm,height=3.5cm]{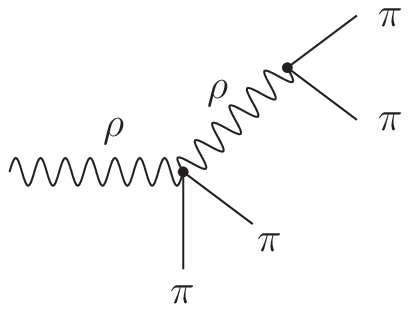}
\includegraphics[width=6cm,height=3.5cm]{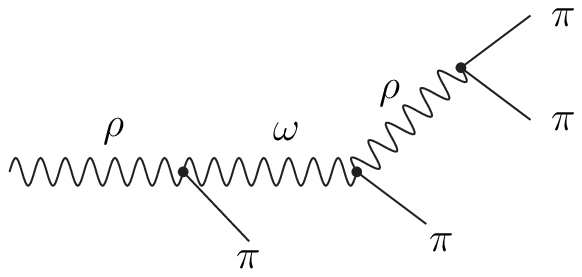}
\end{center}
\vskip -1. cm
\caption{Diagrams contributing to the hadronic current in \cite{Czyz:2000wh}
(upper) and the new contributions from $\rho$ mesons (middle) 
and the omega (lower).}
\label{fun2}
\end{figure}
%%%%%%%%%%%%%%%%%%%%%%%%%%

The latest (still preliminary) accurate
 measurement (by means of the radiative return method) 
 by BaBar \cite{bb1} 
 of the $2\pi^0\pi^+\pi^-$ mode complemented the precise $e^+e^-$ data 
 sets
by CMD2 \cite{Akhmetshin:2004dy}, SND \cite{Achasov:2003bv}  
 and BaBar \cite{Aubert:2005eg}.
Together with the
 $\tau$ data from ALEPH \cite{Schael:2005am} 
and CLEO\cite{Edwards:1999fj} it allows for model independent isospin
 symmetry tests.

 Combining the results from BaBar \cite{Aubert:2005eg,bb1}
  and using relations between $\tau$
and $e^+e^-$ Eq.(\ref{CVC}), we obtain predictions for 
the $\tau$ spectral functions $v$ 
 (related directly to $R^\tau$ \cite{4pi}).
 We use normalization of
the spectral functions chosen by ALEPH \cite{Schael:2005am}.

As shown in Fig.\ref{fun11} and Fig.\ref{fun12} there 
is a good agreement between the spectral functions
 predicted from the BaBar data and the isospin symmetry
 and the ones measured by ALEPH and CLEO. 
One observes systematic shifts which are however
 well contained within current error bars and
  it is not possible to claim an 
observation of the
isospin symmetry violation.

Old $4\pi$ model adopted from \cite{Decker:1987mn}
 and used in \cite{Czyz:2000wh} cannot 
reproduce new and more accurate data. The update to the model 
 has been constructed in \cite{4pi}. Amplitudes
used in the new model are schematically depicted in Fig.\ref{fun2}. 
 The upper diagrams show old contributions from  \cite{Czyz:2000wh},
 the middle diagrams represent newly added contributions
 where the $\rho$ particles are
treated as SU(2) gauge bosons and the lower diagram represent
new omega contributions, which substitute the last diagram
 of the first line.

%\vskip +5cm

There are following parameters in the model: external masses and widths 
$m_{\rho'}$,  $\Gamma_{\rho'}$,$m_{\rho''}$,  $\Gamma_{\rho''}$,
$m_{\rho'''}$,  $\Gamma_{\rho'''}$, 
four couplings in each of the $a_1$, $f_0$
and $\omega$ parts and one coupling in $\rho$-part. 
The parameters were fitted to the existing data. 
The fit is 
quite good, with $\chi^2 / n_{d.o.f} = 275 / 287$.

The comparison between new model and the data from ALEPH, CLOE and
BaBar are shown in Fig.\ref{fun11} and Fig.\ref{fun12}.
The upper and lower curves represent error bars.

It is interesting to compare also the $\tau$ branching ratios from 
the PDG \cite{Yao:2006px}, the new model predictions
 and the direct predictions from BaBar data  
 using isospin symmetry. The results are collected in
 Table \ref{tab1}. They are in agreement within current error bars.
  One can observe about two sigma difference
between PDG and BaBar data for Br($\tau^-\to\nu_\tau2\pi^-\pi^+\pi^0$).
 Better precision BaBar data, expected after the preliminary
 results \cite{bb1} are published, might shed light on the
  isospin violation in the four pion $e^+e^-$ production and $\tau$ decays.

%%%%%%%%%%%%%%% TABLE
\begin{table}
\begin{center}
\begin{tabular}{c|c}
\hline
  & Br($\tau^-\to\nu_\tau2\pi^-\pi^+\pi^0$) \\
\hline
 PDG \cite{Yao:2006px} &(4.46$\pm$ 0.06)\%  \\
\hline
 model &(4.12 $\pm$ 0.21)\%  \\
\hline
 BaBar (CVC) & (3.98 $\pm$ 0.30)\% \\
\hline
\end{tabular}
\end{center}
\end{table}
%
%%%%%%%%%%%%%%% TABLE
\begin{table}
\begin{tabular}{c|c|c}
\hline
  Br($\tau^-\to\nu_\tau\pi^-\omega(\pi^-\pi^+\pi^0)$) & 
Br($\tau-\to\nu_\tau\pi^-3\pi^0$)\\
\hline
(1.77$\pm$ 0.1)\% & (1.04$\pm$ 0.08)\% \\
\hline
(1.60 $\pm$ 0.13)\% & (1.06 $\pm$ 0.09)\% \\
\hline
(1.57 $\pm$ 0.31)\% & (1.02$\pm$ 0.05)\% \\
\hline
\end{tabular}
\vskip 0.3cm
\caption{Branching ratios of $\tau$ decay modes. Comparison between 
 model, experimental data \cite{Yao:2006px}
 and predictions based on BaBar data \cite{Aubert:2005eg,bb1}
 and isospin symmetry.}
\label{tab1}
\end{table}

% SECTION

\section{The narrow resonances J/$\psi$ and $\psi$(2S) in PHOKHARA}

The implementation the narrow resonances  J/$\psi$ and $\psi$(2S)
to the PHOKHARA event generator is the latest update of this
program. The final results will be presented soon and in this paper
we present the preliminary results.

Contributions from two narrow resonances
\\

$J/\psi \ \ \ \to m $ = 3096.916 MeV, $\Gamma$ = 93.4 keV

$\psi(2S) \to  m $ = 3686.093 MeV, $ \Gamma$ = 337 keV,
\\
\\
   to  the final states:
\\

$\pi^+ \pi^-$, $\mu^+ \mu^-$, $K^+ K^-$ and $K^0 \bar{K^0}$

\noindent
 were
 implemented in PHOKHARA.

\begin{figure}[ht]
\begin{center}
\includegraphics[width=4cm,height=3cm]{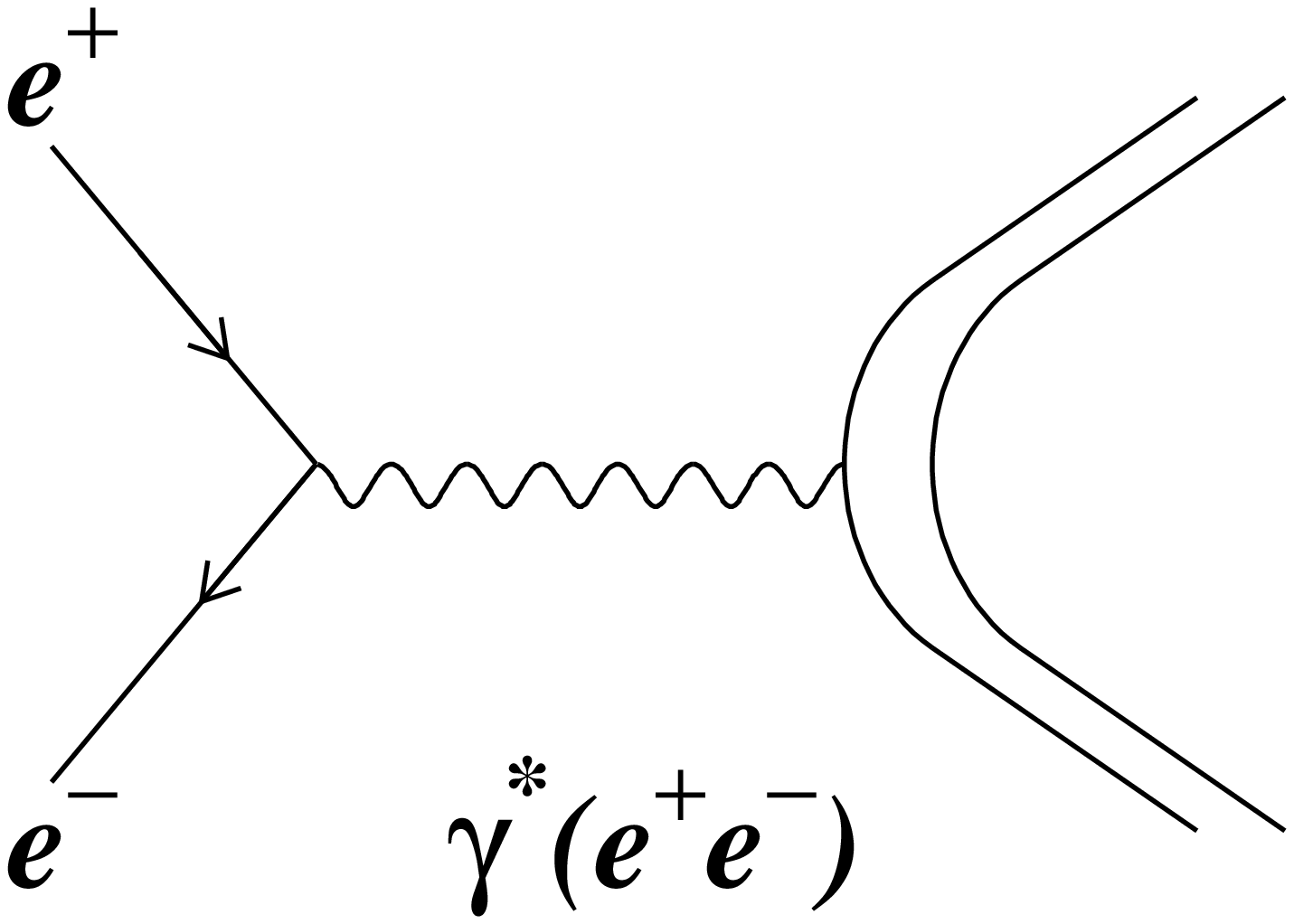}\includegraphics[width=4cm,height=3cm]{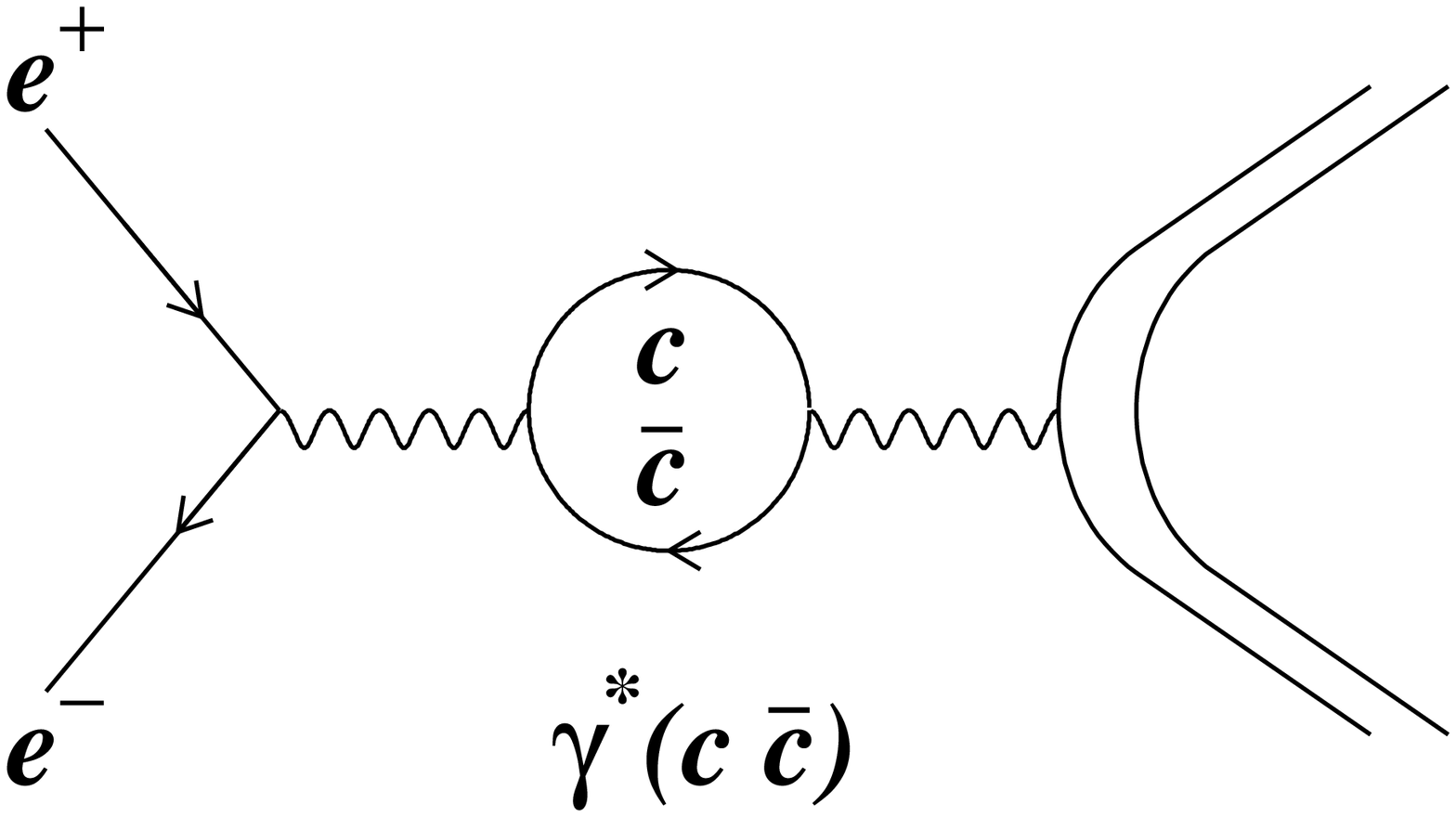}
\includegraphics[width=4cm,height=3cm]{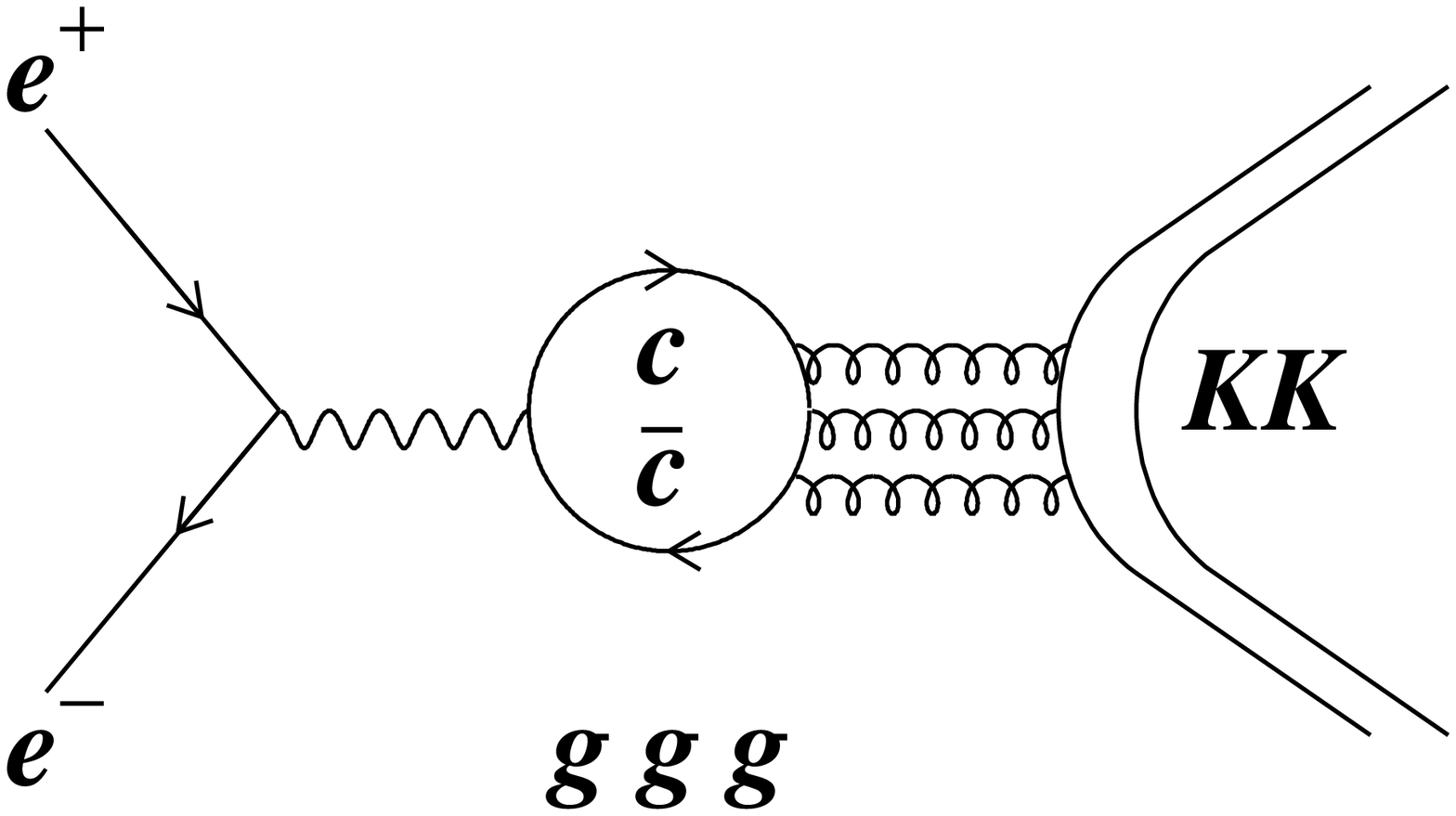}
\end{center}
\vskip -1.3 cm
\caption{The Feynman diagrams of $e^+ e^- \to \mu^+ \mu^-, \pi^+ , \pi^-, KK$
at charmonium resonance,  the one-photon continuum process, the one-photon
decays and the three-gluon decays (only for kaons).}
\label{fun3}
\end{figure}

In general one has three types of contributions to the
 production amplitude at the narrow resonance
(Fig.\ref{fun3}): one-photon continuum, one-photon decays
and three-gluon decays. The last amplitude contributes
only to the  $K^+ K^-$ and $K^0 \bar{K^0}$
production.

%%%%%%%%%%%%%%%%%%%%%%%% 
%%%%%%%%%%%%%%%%%%%%%%%%

For reliable predictions of the $\pi^+ \pi^-$ and $KK$ production
  good models of  electromagnetic pion and kaon form factors are needed.
The form factors used in the public version of PHOKHARA 6.0 were
  taken from \cite{Bruch}. The CLEO-c measurement \cite{Pedlar:2005sj}
  and results of the investigations in 
 \cite{Milana:1993wk} and \cite{Seth} were not accounted for in
  \cite{Bruch}. As a result,
  the values of form factors taken from \cite{Bruch}
  at the $J/\psi$ and $\psi(2S)$ are
  much smaller then the ones obtained in 
\cite{Pedlar:2005sj,Milana:1993wk,Seth} and new investigations were necessary.

 For the pion form factor we keep the structure of the model
 from \cite{Bruch}:

\begin{eqnarray}
&& F_\pi(s)= \left[\sum\limits_{n=0}^N c_{\rho_n} BW_n(s)\right]_{fit}\nonumber \\
&&\kern+27pt 
+ \left[\sum\limits_{n=N+1}^{\infty}c_{\rho_n} BW_n(s)\right]_{dual-QCD_{N_c=\infty}}\,,
\nonumber \\
\label{themodel}
\end{eqnarray}
where firsts N-1 couplings are fitted ($c_N$
is calculated). We perform the fit taking into account the experimental data
 not accounted for in \cite{Bruch}.
  For kaons, unlike in \cite{Bruch}, 
 we use infinite towers of resonances
 for $\rho$, $\omega$ and $\phi$ (see \cite{narrow} for more details).
%
%%%%%%%%%%%%%%%%%%%%%%%%
 \begin{figure}[ht]
\includegraphics[width=7.5cm,height=6cm]{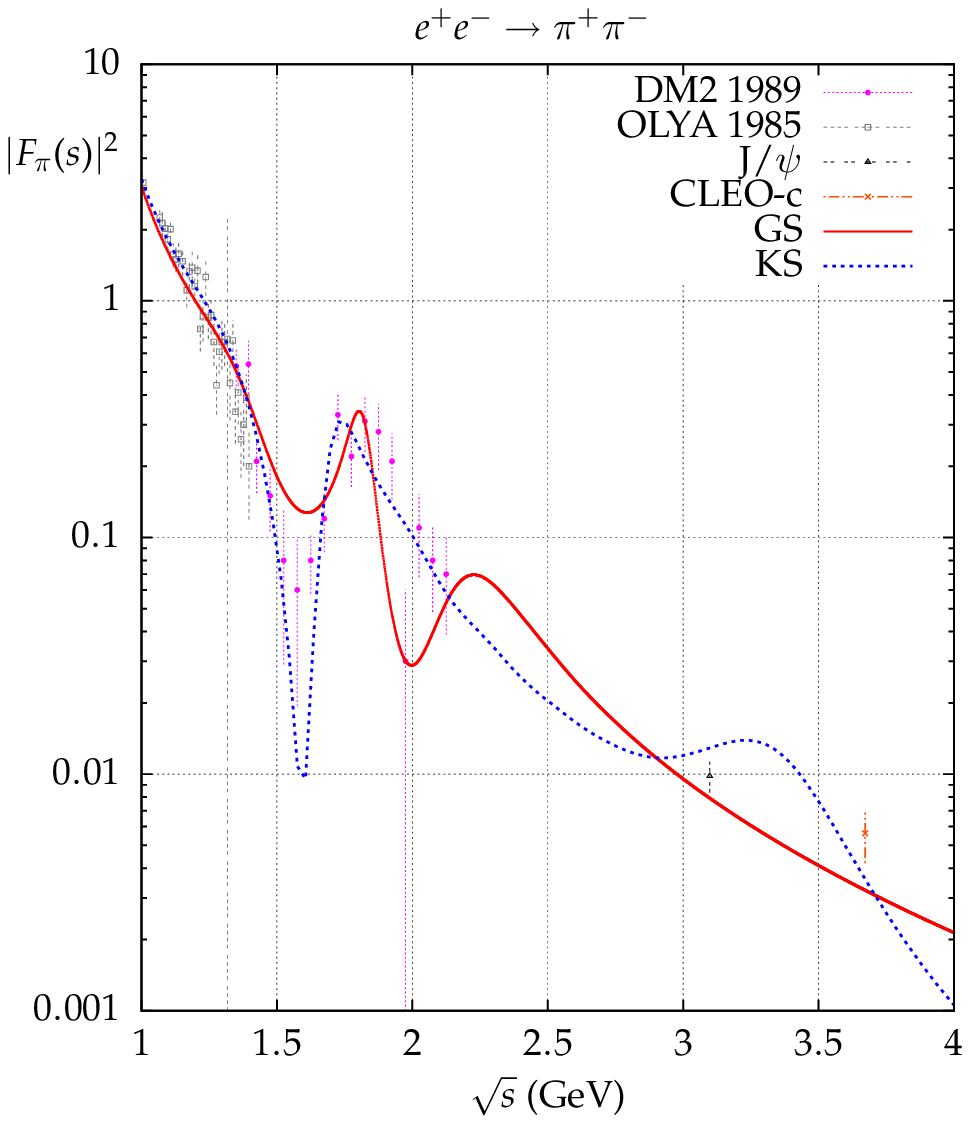}
\includegraphics[width=7.5cm,height=6cm]{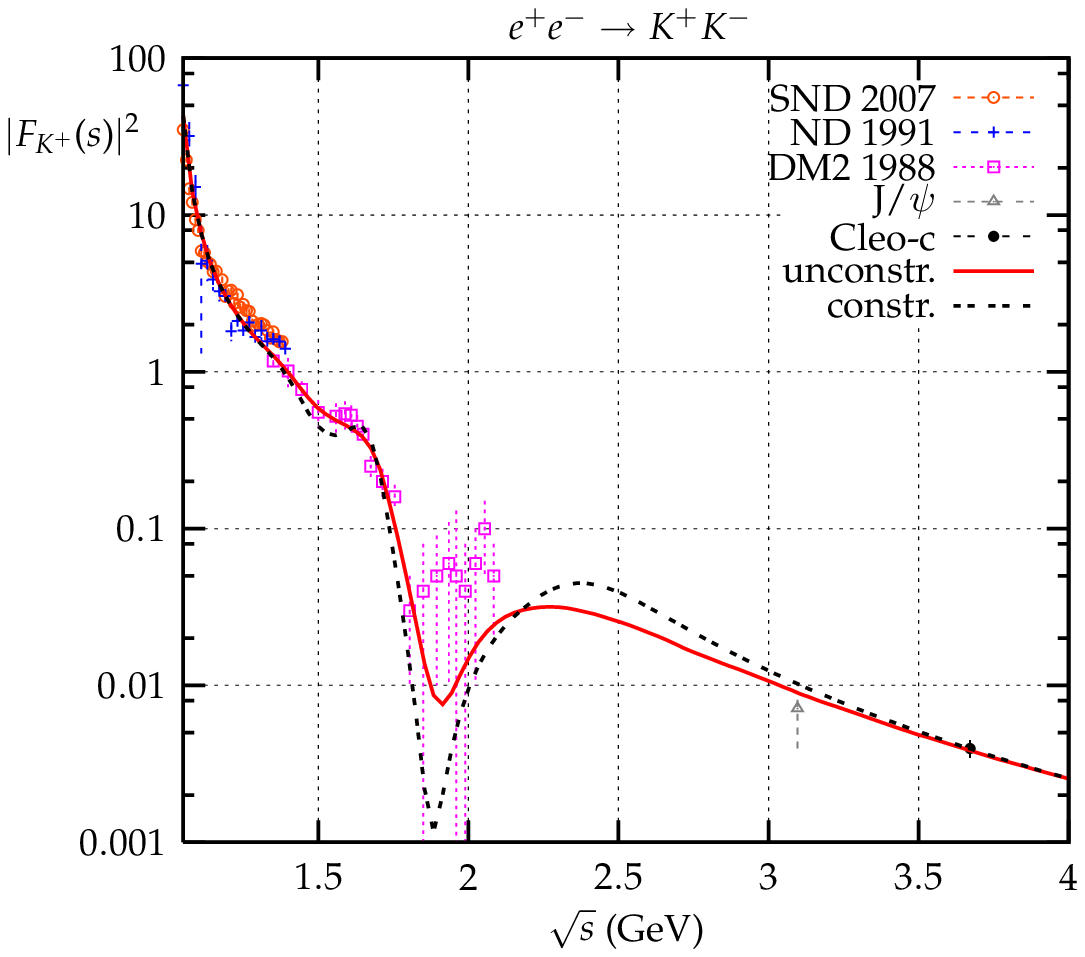}
\vskip -1. cm
\caption{The experimental data \cite{Pedlar:2005sj,Bisello:1988hq} 
compared to the model fits
results. The form factor at J/$\psi$ is from \cite{Seth}
(theoretical extraction).}
\label{FF}
\end{figure}
%%%%%%%%%%%%%%%%%%%%%%%%%
%
The results of the fits are summarized in
Fig.\ref{FF}.
For pions we considered two versions of the model,
where the Breit-Wigner function was taken at tree level 
\cite{KS} - K\"uhn-SantaMaria (KS) model and where it was taken
with pion loop corrections \cite{GS} - Gounaris-Sakurai (GS) model.
For kaons we also considered two versions of the model,
constrained and unconstrained (see \cite{Bruch} for the details).

%
%%%%%%%%%%%%%%%%%%%%%%%%%
 \begin{figure}[ht]
\includegraphics[width=7.5cm,height=6cm]{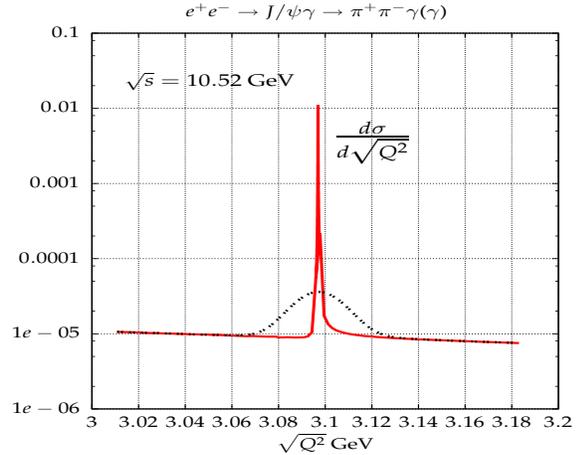}
\caption{Differential cross section for $\sqrt{s}$ = 10.52 GeV
of the process $e^+e^- \to J/\psi \gamma \to \pi^+\pi^- \gamma (\gamma)$
without (solid line) and with the detector smearing effect.}
\label{fun4}
\end{figure}
%%%%%%%%%%%%%%%%%%%%%%%%%

One has to remember that due to
the finite detector resolution one never observes the true
distribution of the events but its convolution with the detector
 resolution function. This is extremely important for studies of
 the narrow resonances, where typical energy resolution is much bigger
 than a width of a resonance.
 Fig.\ref{fun4} shows the differential cross section of the 
process   $e^+e^- \to J/\psi \gamma \to \pi^+\pi^- \gamma (\gamma)$
with true distribution of the events (solid line) and 
the realistic differential cross section obtained by 
 smearing with the Gaussian distribution 
 with the 14.5 MeV standard deviation (taken from \cite{detector}).

\begin{figure}[ht]
\includegraphics[width=7.5cm,height=6cm]{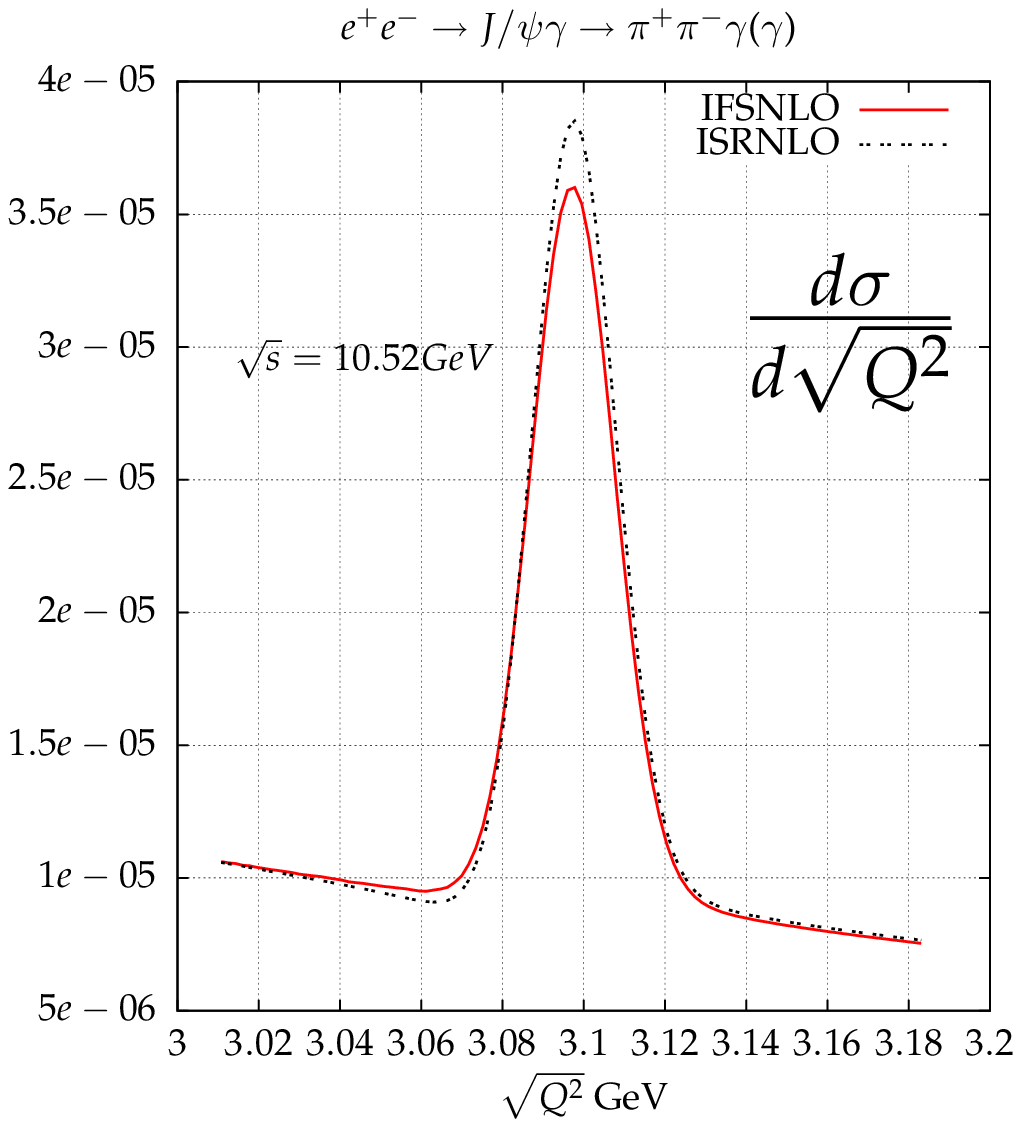}a
\includegraphics[width=7.5cm,height=6cm]{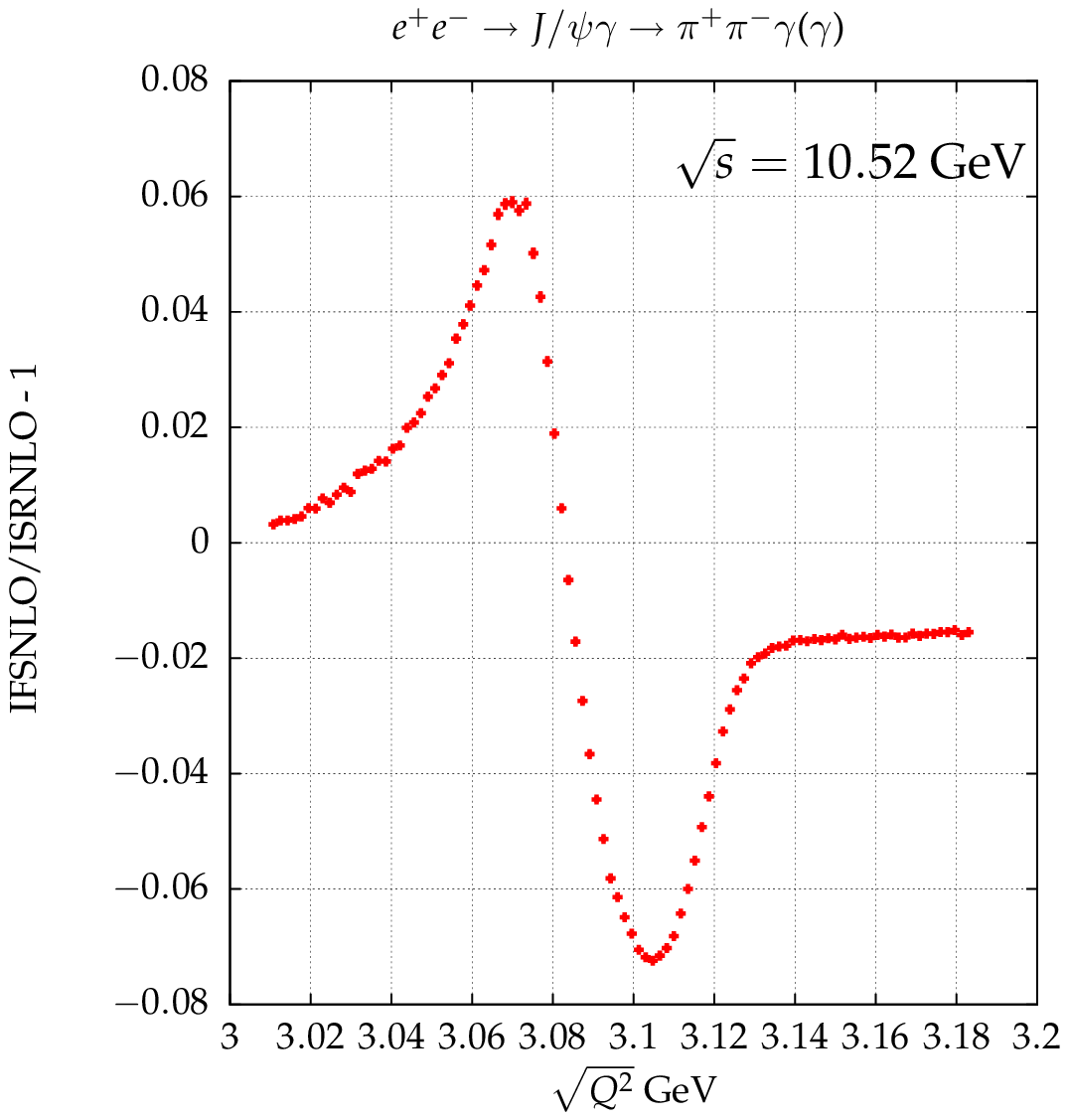}
\vskip -1. cm
\caption{The size of the next-to-leading FSR correction (IFSNLO)
compared with the next-to-leading ISR correction (ISRNLO).
Detector smearing effect are taken into account.}
\label{fun5}
\end{figure}

We investigated also the role of the FSR radiation (at next to leading order)
 on the radiative
return cross sections in the vicinity of the narrow resonances.
In Fig.\ref{fun5} we show the results for two pion final state. 
 Relatively big corrections to the differential
 $Q^2$ distribution, coming from FSR (IFSNLO=FSRNLO+ISRNLO) are
 observed as compared to the ISRNLO only, even if the integrated cross sections
 differ only by about 2\%.

%for pions $\sigma_{IFSNLO}$ = 2.27808(13) fb, $\sigma_{ISRNLO}$ = 2.32720(6) fb
%for muons  $\sigma_{IFSNLO}$ = 6.8519(8) pb, $\sigma_{ISRNLO}$ = 6.7986(8) pb

%%%%%%%%%%%%%%%%%%%%%%%%%
%%%%%%%%%%%%%%%%%%%%%%%%%

\section{Conclusions}

The present status of the PHOKHARA event generator was described.
The four-pion channels reanalysis and the latest
upgrade of the generator, implementation of 
 narrow resonances, were presented.

\end{document}